\documentclass[pra,twocolumn,superscriptaddress,showpacs,floatfix,amssymb,
amsmath,fontsize=12pt]{revtex4}

\usepackage{times}
\usepackage{amssymb}
\usepackage{graphicx}
\usepackage{dcolumn}
\usepackage[latin1]{inputenc}
\usepackage[mathscr]{eucal}
\usepackage{epsfig}
\usepackage{rotating}
\usepackage{bm}
\usepackage{bbm}
\usepackage{textcomp}
\usepackage{sidecap}
\usepackage{times}
\usepackage[american]{babel}

\newcommand{\be}{\begin{equation}}
\newcommand{\ee}{\end{equation}}
\newcommand{\bee}{\begin{equation*}}
\newcommand{\eee}{\end{equation*}}
\newcommand{\br}{\bm{r}}

\newcommand{\bk}{\bm{k}}
\newcommand{\la}{\langle}
\newcommand{\ra}{\rangle}
\newcommand{\al}{\alpha}

\newcommand{\da}{\dagger}

\newcommand{\dr}{\mathrm{d}\br}

\newcommand{\vtr}{V_{\mathrm{tr}}^{}}

\newcommand{\bff}{\bm{\mathcal{F}}}

\newcommand{\hz}{\hat{h}_0^{}}

\newcommand{\fiis}{\hat{\phi}}
\newcommand{\hn}{\hat{n}}
\newcommand{\hc}{H_{\scriptscriptstyle\bm{C}}}
\newcommand{\nc}{N_{\scriptscriptstyle\bm{C}}}
\newcommand{\kb}{k_{\mathrm{B}}^{}}

\newcommand{\psiic}{\vec{\Psi}_{{\scriptscriptstyle\bm{C}}}^{}}
\newcommand{\psiicd}{\vec{\Psi}_{{\scriptscriptstyle\bm{C}}}^{\da}}
\newcommand{\psic}{\psi_{{\scriptscriptstyle\bm{C}},\alpha}^{}}

\newcommand{\md}{\mathcal{D}}
\newcommand{\mdp}{\mathcal{D}_{\scriptscriptstyle P}^{}}
\newcommand{\mdq}{\mathcal{D}_{\scriptscriptstyle Q}^{}}

\newcommand{\ecut}{\varepsilon_{\mathrm{cut}}^{}}

\newcommand{\nlsm}{NL$\sigma$M}

\begin{document}

\title{Finite-temperature phase transitions in quasi-two-dimensional spin-$1$ 
Bose gases}

\author{Ville Pietil\"a}
\affiliation{Department of Applied Physics/COMP, Aalto
  University, P.~O.~Box 15100, FI-00076 AALTO, Finland}
\affiliation{Australian Research Council Centre of Excellence for Quantum Computer Technology, School of Electrical Engineering \& Telecommunications,
University of New South Wales, Sydney NSW 2052, Australia}
\author{Tapio P.~Simula}
\affiliation{Department of Physics, Okayama University, Okayama 700-8530, 
Japan}
\affiliation{School of Physics, Monash University, Victoria 3800, Australia}
\author{Mikko M\"ott\"onen}
\affiliation{Department of Applied Physics/COMP, Aalto
  University, P.~O.~Box 15100, FI-00076 AALTO, Finland}
\affiliation{Australian Research Council Centre of Excellence for Quantum Computer Technology, School of Electrical Engineering \& Telecommunications,
University of New South Wales, Sydney NSW 2052, Australia} 
\affiliation{Low Temperature Laboratory,
Aalto University, P.~O.~Box 13500, FI-00076 AALTO,
Finland}

\begin{abstract}
Recently, the Berezinskii-Kosterlitz-Thouless transition was found to be 
mediated by half-quantum vortices (HQVs) in two-dimensional (2D) 
antiferromagnetic Bose gases [Phys.~Rev.~Lett.~{\bf 97}, 120406 (2006)]. We 
study the thermal activation of HQVs in the experimentally relevant trapped 
quasi-2D system and find a crossover temperature at which free HQVs 
proliferate at the center of the trap. Above the crossover 
temperature, we observe transitions corresponding to the onset of a coherent 
condensate and a quasicondensate, and discuss the absence of a fragmented 
condensate.
\end{abstract} 


\maketitle

\section{Introduction} 
The dimension of the underlying space has a profound 
impact on the existence of long-range order and phase transitions in a given  
system. In two-dimensional (2D) systems the long-range order and spontaneous 
symmetry breaking are forbidden~\cite{Mermin:1966,Hohenberg:1967,
Coleman:1973}; however, 2D systems can exhibit a quasi-long-range order with 
algebraically decaying correlations~\cite{Berezinskii:1971,Berezinskii:1972,
Kosterlitz:1972,Kosterlitz:1973}. The disordered high-temperature phase and 
the algebraically ordered low-temperature phase are separated by a topological 
phase transition corresponding to the unbinding of pairs of vortices and 
antivortices. This phase transition is referred to as the 
Berezinskii-Kosterlitz-Thouless (BKT) transition~\cite{Berezinskii:1971,
Kosterlitz:1972}. However, experimentally relevant examples often display 
additional features due to the finite-size effects~\cite{Bramwell:1993}, and 
in the trapped ultracold atomic gases where the BKT transition has recently 
been studied~\cite{Simula:2006,Hadzibabic:2006,Schweikhard:2007,Holzmann:2008,
Clade:2009}, the inhomogeneous density of the gas renders the superfluid 
state and the coherence properties qualitatively different from those of 
the bulk systems~\cite{Clade:2009,Simula:2008,Bisset:2009}.

Spinor Bose gases~\cite{Ohmi:1998,Ho:1998,Stenger:1998} are especially 
interesting as they can in principle combine magnetic ordering, formation 
of a condensed component, and superfluidity. Due to the interplay of 
these competing orders, the antiferromagnetic spin-$1$ Bose gas is expected to 
host various exotic phenomena such as fragmented condensates~\cite{Ho:2000} 
and fractionalized topological objects~\cite{Leonhardt:2000,Demler:2002} that 
are usually absent in the single-component systems. For example, a 
half-quantum vortex (HQV) confined to a spin defect occurs in spin nematic 
condensates~\cite{Leonhardt:2000,Ji:2008} and it has recently been 
created using Raman-detuned laser pulses~\cite{Wright:2009}. 
In homogeneous 2D optical lattices, proliferation of HQVs in spin-$1$ Bose 
systems due to thermal fluctuations has been predicted~\cite{Podolsky:2009,
Song:2009}, and the superfluid transition in two dimensions was found to be 
mediated by HQVs~\cite{Mukerjee:2006}. In ferromagnetic spinor condensates, 
HQVs have shown to give rise to intriguing magnetization 
dynamics~\cite{Saito:2008}. Fractional vortices and the related BKT 
transitions have also been discussed in the context of 
$^3$He~\cite{Salomaa:1985} and different nonconventional 
superconductors~\cite{Babaev:2004,Babaev:2005,Chung:2007}. Recently, HQVs 
have been observed in exciton-polariton condensates~\cite{Lagoudakis:2009}.

While the connection between superfluidity and Bose-Einstein 
condensation is relatively well understood in the single-component Bose 
systems~\cite{Berezinskii:1971,Kosterlitz:1972,Simula:2006,Hadzibabic:2006,
Schweikhard:2007,Holzmann:2008,Clade:2009,Simula:2008,Bisset:2009}, the 
existence of spin degree of freedom in spinor Bose gases renders the relation 
between superfluidity and long-range order more complicated and far less 
studied. In particular, the existence and the nature of the possible condensed 
component is not yet known in two dimensions.
The recent experimental interest in spinor Bose gases with 
antiferromagnetic interactions~\cite{Liu:2009,Liu:2009b}, advances 
in the evaporative cooling of optically trapped atoms~\cite{Hung:2008}, and 
the nondestructive imaging of the local magnetization of spin-$1$ Bose 
gases~\cite{Carusotto:2004,Higbie:2005} suggest that the experimental 
realization of the finite-temperature phase transitions in quasi-2D spinor 
Bose gases may be possible in the near future. Hence, we study the activation 
of different topological defects associated with the superfluid transition and 
determine the different degenerate components of quasi-2D antiferromagnetic 
spin-$1$ Bose gases. Our approach is valid in the regime where the thermal 
fluctuations are dominant and our results suggest that in this region, the 
condensate state is nonfragmented.

\section{Formalism\label{formalism}} 
To study the behavior of a spinor Bose gas near the 
critical region we use a classical field (c field) to describe the 
highly occupied low-energy modes and a quantum field for the thermal 
modes with low occupation~\cite{Blakie:2008}. Previously, this approach has 
been successfully applied in studies of the BKT transition in scalar Bose 
gases~\cite{Simula:2006,Simula:2008,Bisset:2009,Bisset:2009b} as well as 
to predict other properties of dilute scalar Bose 
gases~\cite{Davis:2001,Davis:2003,Davis:2006,Davis:2007,Bezett:2008,
Bezett:2009,Sato:2009}. The coherence properties of spinor Bose condensates 
at finite temperatures have recently been studied using an alternative 
formulation of the c-field method~\cite{Gawryluk:2007}.

The dynamics of the c field can be described by the projected 
Gross-Pitaevskii equation (PGPE)
\begin{equation}
\label{pgpe} i\hbar\,\partial_t\psiic
=\hat{h}_0\psiic + \mathcal{P}\big\{c_0^{}|\psiic|^2\psiic +
c_2^{}(\psiicd\bm{\mathcal{F}}\psiic)\cdot\bm{\mathcal{F}}\psiic\big\},
\end{equation} 
which generalizes the usual spinor Gross-Pitaevskii 
equation~\cite{Ohmi:1998,Ho:1998}. In the PGPE, $\bm{\mathcal{F}}$ denotes a 
vector of spin-$1$ matrices and $\mathcal{P}$ implements a projection into 
the subspace of the classical modes~\cite{Blakie:2008}. The c field in 
Eq.~\eqref{pgpe} is written in the basis consisting of the Zeeman substates 
such that $\psiic = (\psic)$, $\al=1,0,-1$. The single-particle operator 
$\hat{h}_0$ is given by 
\begin{equation*}
\hat{h}_0=-\frac{\hbar^2}{2m}\nabla^2 + 
\frac{m}{2}(\omega_{\perp}^2r^2_{\perp}+\omega_z^2 z^2).
\end{equation*}
The coupling constants $c_0^{}$ and $c_2^{}$ are given by $c_0^{} = 
4\pi\hbar^2(a_0^{}+2a_2^{})/3m$ and $c_2^{} = 
4\pi\hbar^2(a_2^{}-a_0^{})/3m$, where $m$ is the atomic mass and $a_0^{}$ and 
$a_2^{}$ are the $s$-wave scattering lengths in the total hyperfine spin 
channels $F=0$ and $F=2$, respectively~\cite{Ho:1998}. Antiferromagnetic 
interactions imply $c_2^{} > 0$, and we take 
$a_0^{}=46a_B^{}$, $a_2^{}= 52a_B^{}$, and $m=3.86\times 10^{-26}$ kg, 
according to $^{23}$Na~\cite{Ho:1998}. The Bohr radius is denoted by $a_B^{}$.

In the quasi-2D limit,  $\omega_{\perp}^{} \ll \omega_z^{}$ and we choose 
$\omega_z^{}=200\times\omega_{\perp}^{}$. Harmonic oscillator lengths 
in axial and transverse directions are denoted by 
$a_{z}^{}=\sqrt{\hbar/m\omega_{z}}$ and 
$a_{\perp}^{}=\sqrt{\hbar/m\omega_{\perp}}$. The scattering can be treated as 
three-dimensional as long as $a_0^{},a_2^{} \ll a_z^{}$~\cite{Petrov:2000,
Bloch:2008}. This condition is satisfied with the previous 
choice of $\omega_z^{}$ and $\omega_{\perp}^{}$ if we take $\omega_{\perp} = 
2\pi\times 10$ Hz which is in the realm of the current experiments.
Since we consider a quasi-2D situation, the PGPE and all c fields 
are defined in a 3D space. The spatial vector of the 3D space is denoted by 
$\br$ and $\br_{\perp}^{}$ is a 2D vector in the $x-y$ plane. 
Summation over repeated indices is implied. 

In the PGPE, the c-field region $\bm{C}$ is defined by the energy cutoff 
$\ecut$ such that $\bm{C} = \{ n\,|\,\varepsilon_n^{} \leq \ecut\}$,  
corresponding to the spectrum of the single-particle operator $\hat{h}_0^{}$.
The c fields in Eq.~\eqref{pgpe}  can be expressed in terms of the 
eigenstates of  $\hat{h}_0^{}$ 
\be
\label{c-field}
\psic(\br) = \sum_{n\in\bm{C}} c_{\al,n}^{}\varphi_n^{}(\br).
\ee  
The PGPE corresponds to a microcanonical system in which the stationary 
probability distributions are determined by the total energy of system, and 
the temperature and the chemical potential are computed as ensemble averages.
We use the ergodic hypothesis to replace all ensemble averages with the 
corresponding time averages. Using the ergodic hypothesis, thermodynamical 
quantities such as the temperature and the chemical potential can be computed 
dynamically~\cite{Blakie:2008,Davis:2003,Davis:2005}. 

Let us briefly discuss how to generalize the single-component calculation of 
the temperature and the chemical potential~\cite{Blakie:2008,Davis:2003,
Davis:2005} to the spin-$1$ case. The PGPE~\eqref{pgpe}  arises from the 
Hamiltonian 
\be
\hc = \int\dr\,\big[\psiicd\hz\psiic + \frac{c_0^{}}{2}|\psiic|^4 +
\frac{c_2^{}}{2}(\psiicd\bff\psiic)^2\big]
\ee
for which the canonical 
coordinates can be defined such that 
\begin{equation}
\label{can_coord}
Q_{\alpha,n}^{} = \frac{1}{\sqrt{2\varepsilon_n^{}}}(c_{\alpha,n}^{*} + 
c_{\alpha,n}^{})\,\,\,\mathrm{and}\,\,\,
P_{\alpha,n}^{} = i\sqrt{\frac{\varepsilon_n^{}}{2}}(c_{\alpha,n}^{*} - 
c_{\alpha,n}^{}),
\end{equation}
where $c_{\alpha,n}^{}$ are the coefficients in Eq.~\eqref{c-field}.
The canonical coordinates are collectively denoted by 
$\bm{\Gamma} = \{Q_{\alpha,n}^{},P_{\alpha,n}^{}\}$. 
According to a general theorem~\cite{Rugh:1997}, the temperature can be 
calculated as
\be
\label{temperature}
\frac{1}{\kb T} \equiv \left( \frac{\partial S}{\partial E}\right)_N = 
\la \mathcal{D}\cdot\bm{X}_T^{}(\bm{\Gamma})\ra,
\ee
where the first identity is the standard definition of the temperature of a 
microcanonical system. The derivative operator 
$\mathcal{D} = \{e_n^{}\partial/\partial_{\Gamma_n^{}}^{}$\}, determined by 
the coefficients $\{e_n^{}\}$,  and the vector field 
$\bm{X}_T^{}$ can be chosen freely as long as they satisfy the 
conditions~\cite{Davis:2003,Davis:2005,Blakie:2008}
\be
\label{t_cond}
\md\hc\,\cdot\,\bm{X}_T^{} = 1\,\,\,\mathrm{and}
\,\,\,
\md\nc\,\cdot\,\bm{X}_T^{} = 0,
\ee
where $\nc = \int\dr\,|\psiic|^2$ is the total number of the c-field atoms. 
The vector field $\bm{X}_T^{}$ satisfying the above constraints is given 
by~\cite{Davis:2003,Davis:2005} 
\be
\label{vector_field}
\bm{X}_T^{} = \frac{\md\hc - \lambda_N^{}\md\nc}{|\md\hc|^2 - \lambda_N^{}
(\md\nc\,\cdot\,\md\hc)},
\ee
with $\lambda_N^{} = (\md\nc\,\cdot\,\md\hc)/|\md\nc|^2$. 
Straightforward choices for the vector operator $\md$ are 
$\mdp = \{0,\partial_{{\scriptscriptstyle P}_n^{}}^{}\}$ and 
$\mdq = \{\partial_{{\scriptscriptstyle Q}_n^{}}^{},0\}$~\cite{Davis:2005}. 
The temperature is independent of the choice the derivative 
$\md$, and the two different choices serve also as a check for the numerical 
implementation. The average in Eq.~\eqref{temperature} is computed as a 
corresponding time average.

The preceding formulation can be applied when the only conserved quantity is 
the total particle number $\nc$. In the present case, also the angular 
momentum is conserved. Transforming to the coordinate system with zero 
total angular momentum, the angular momentum conservation does not appear in 
Eqs.~\eqref{t_cond} and~\eqref{vector_field}~\cite{Rugh:2001,Davis:2005}. 
For spinor Bose gases, the conservation of the total magnetization also needs  
to be taken into account. In the antiferromagnetic case, however, the total 
magnetization is zero and can be neglected in light of the previous 
argument. Using the definition $\mu/\kb T = - (\partial S/\partial N)_E^{}$, 
also the chemical potential $\mu$ can be computed by interchanging the roles 
of $\hc$ and $\nc$. Computationally efficient formulation for the different 
terms in Eqs.~\eqref{temperature} -- \eqref{vector_field} proceeds in an 
analogous way to Ref.~\cite{Davis:2005}.

The number of atoms outside the c-field region can be computed 
self-consistently using the Hartree-Fock-Popov (HFP) 
approximation~\cite{Simula:2006,Blakie:2008,Simula:2008,Zhang:2004}. 
The full field operator containing the c-field part $\psic$ and 
the incoherent part $\delta\hat{\phi}_{I,\alpha}$ is denoted by 
$\hat{\Phi}_{\alpha}^{} = \psic + \delta\hat{\phi}_{I,\alpha}^{}$.  
We assume that terms such as 
$\langle \psic\delta\hat{\phi}_{I,\beta}^{}\rangle$, 
$\langle \psic\delta\hat{\phi}_{I,\beta}^{\dagger}\rangle$,  
and all their complex conjugates vanish.
This leads to the HFP single-particle energies~\cite{Zhang:2004}
\begin{subequations}
\label{hfp_energies}
\begin{align}
&\varepsilon_+^{}(\bk,\br) = \frac{\hbar^2\bk^2}{2m} + \vtr(\br) -\mu 
+ c_0^{}(n+n_+^{})  \notag \\ 
& \hspace{5cm} +c_2^{}(2n_+^{} + n_0^{} - n_-^{}), \\
&\varepsilon_0^{}(\bk,\br) = \frac{\hbar^2\bk^2}{2m} + \vtr(\br) -\mu 
+ c_0^{}(n+n_0^{}) + c_2^{}(n_+^{} +  n_-^{}), \\
&\varepsilon_-^{}(\bk,\br) = \frac{\hbar^2\bk^2}{2m} + \vtr(\br) -\mu 
+ c_0^{}(n+n_-^{}) \notag \\
& \hspace{5cm}  +c_2^{}(2n_-^{} + n_0^{} - n_+^{}),
\end{align}
\end{subequations} 
where  $n_{\al}^{} = \la \hat{n}_{\al}^{}\ra = |\psic|^2 + 
\la\delta\fiis_{I,\al}^{\da}\delta\fiis_{I,\al}^{}\ra$ and 
$n = \la\hn\ra = n_+^{} + n_0^{} + n_-^{}$.
The occupation number $n_{\al}^{\scriptscriptstyle(I)} = 
\la\delta\fiis_{I,\al}^{\da}\delta \fiis_{I,\al}^{}\ra$ for the incoherent 
atoms can be computed from the Bose-Einstein distribution using the 
semiclassical integral in a 3D phase space
\be
\label{be_distribution}
n_{\al}^{\scriptscriptstyle (I)}(\br) = \int\frac{\mathrm{d}\bk}{(2\pi)^3}\,\,
\frac{1}{e^{\varepsilon_{\al}^{}(\bk,\br)/\kb T}-1}.
\ee 
The quasi-2D nature of the system is taken into account  by treating the axial 
modes discretely in the semiclassical integral~\cite{Simula:2008}. The energy 
cutoff $\ecut$ introduces a spatially dependent low-energy cutoff to the 
semiclassical integral~\eqref{be_distribution} (see Ref.~\cite{Simula:2008}). 

\section{Topological defects and nematic order\label{defects}}  
The c field in Eq.~\eqref{pgpe} is written in the 
$z$-quantized basis $\vec{\Psi} = (\psi_{\alpha})$, $\alpha=-1,0,1$, but   
the nematic properties of antiferromagnetic Bose gases are more conveniently 
expressed in the Cartesian representation~\cite{Ohmi:1998,Mueller:2004} 
$\vec{\Psi} = (\psi_{a})$, $a=x,y,z$. The transformation is given by 
$\psi_x^{} = (\psi_1^{} - \psi_{-1}^{})/\sqrt{2}$, 
$\psi_y^{} = i(\psi_1^{} + \psi_{-1}^{})/\sqrt{2}$, and 
$\psi_z^{} = \psi_0^{}$ and the nematic order is described by the spin 
quadrupole moment~\cite{Mueller:2004}
$\mathcal{Q}^{(s)}_{ab} = (\psi_a^*\psi_b^{} + \psi_b^*\psi_a^{} )/ 
(2|\vec{\Psi}|^2)$.
In general, $\mathcal{Q}^{(s)}$ has three distinct nonzero eigenvalues 
and the local magnetic axis $\hat{\bm{n}}$ is defined as the eigenvector 
associated with the largest eigenvalue. 
For $\psi_z^{}\equiv 0$, the magnetic axis is confined into the $x-y$ plane 
and we refer to such case as the ``in-plane'' nematic. 

\begin{figure}[h!]
\centering
\includegraphics[width=0.475\textwidth]{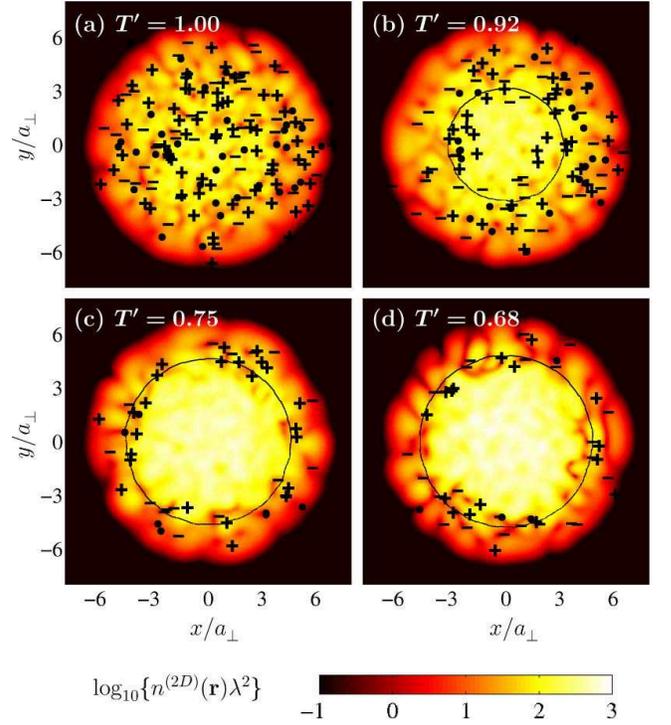}
\caption{\label{densities} (Color online) Instantaneous density of the c-field 
atoms corresponding to the out-of-plane nematic phase. 
Half-quantum vortices and antivortices are denoted by black  $+$ and $-$ 
symbols, respectively. Skyrmions are marked with black dots. 
The black line denotes the boundary outside which  
$n_{c}^{}/\bar{n}_{tot}^{} < 0.1$, where $n_{c}^{}$ is the condensate 
density, and $\bar{n}_{tot}^{}$ is the average total density of the c-field 
atoms. The thermal wavelength is denoted by $\lambda$. In the 
instantaneous density, the $z$-dependence is integrated out. The temperature 
is given with respect to the critical temperature of the corresponding 
quasi-2D ideal Bose gas (see Section~III).}
\end{figure}

In the polar phase which corresponds to identically vanishing local 
spin~\cite{Ho:1998}, the Cartesian representation gives 
$\vec{\Psi} = \sqrt{\varrho}e^{i\theta}\hat{\bm{n}}$ and the HQV corresponds 
to a defect where both $\theta$ and $\hat{\bm{n}}$ have a $\pi$ 
winding about the core of the defect. Furthermore, the polar phase allows also 
the existence of skyrmions which have finite energy and are characterized by 
the second homotopy group of the order parameter space. In the Appendix,  
explicit  expressions for the HQVs and skyrmions are presented. The polar 
phase has a local $\mathbbm{Z}_2^{}$ invariance corresponding to 
$(\theta,\hat{\bm{n}})\rightarrow (\theta+\pi,-\hat{\bm{n}})$~\cite{Zhou:2001,
Demler:2002,Ji:2008}. This implies that defects with opposite topological 
charges cannot be distinguished and therefore we define the sign of the 
HQV from the polarization of the vortex core~\cite{Ji:2008}. Furthermore, we 
do not distinguish between skyrmions with opposite winding 
numbers~\cite{Zhou:2003}. An example of the thermally activated HQVs and 
skyrmions is shown in Fig.~\ref{densities} where the instantaneous 
$z$-integrated densities of the c-field atoms at different temperatures are 
depicted.

\begin{figure} 
\centering
\includegraphics[width=0.475\textwidth]{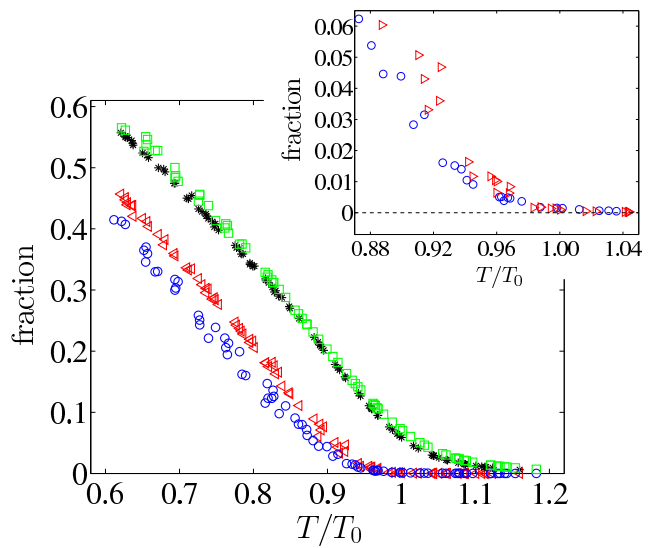}
\caption{\label{condensation} (Color online) Condensate fraction  
$N_0^{}/N_{\mathrm{tot}}^{}$ for the 
in-plane nematic (blue circles) and the out-of-plane nematic (red triangles)
phases as a function of the reduced temperature. The quasicondensate fraction 
is almost identical for the in-plane nematic (green squares) and out-of-plane 
nematic (black asterisks) phases. Note that the definition of the reduced 
temperature differs between the in-plane and the out-of-plane phases (see text 
for details). (Inset) The condensate fraction near the temperature 
corresponding to the onset of the condensate. The dashed line is a guide for 
the eye.}
\end{figure}

We consider two phases, the in-plane nematic with 
$\bar{n}_{c,+1}^{}= \bar{n}_{c,-1}^{}$ and $\bar{n}_{c,0}^{}\equiv 0$ and 
an ``out-of-plane'' nematic with $\bar{n}_{c,+1}^{} \approx \bar{n}_{c,0}^{} 
\approx \bar{n}_{c,-1}^{} = 0.33 \pm 0.06$. 
Here $\bar{n}_{c,\alpha}^{}$ refers to the average number of c-field atoms 
in the component $\alpha$, divided by the total number of atoms 
$\nc$ in the c-field region. The average numbers $\bar{n}_{c,\alpha}^{}$ 
corresponding to the different data points in Figs.~\ref{condensation} 
and~\ref{coh_cond} vary between the aforementioned limits.
In both cases, we take $\nc=15000$ and choose the energy cutoff as 
$\epsilon_{\mathrm{cut}}^{}=126\,\hbar\omega_{\perp}^{}$ 
($\epsilon_{\mathrm{cut}}^{}=122\,\hbar\omega_{\perp}^{}$) for the in-plane 
(out-of-plane) nematic. For these values of the cutoff energy, only the 
lowest axial mode becomes populated and the c-field region is in the  
ground state with respect to motion in the $z$ direction. Nevertheless, the 
incoherent region atoms typically occupy several axial modes.

As indicated in Ref.~\cite{Podolsky:2009}, the 
in-plane nematic phase arises in the spin-$1$ case as a result of a large 
negative quadratic Zeeman shift (for a discussion how the negative shift 
is physically realized, see Refs.~\cite{Santos:2007,Podolsky:2009}). 
In the case of the PGPE, elimination of the $\al=0$ component corresponds to 
leaving the $\al=0$ component empty in the initial state. The quadratic Zeeman 
shift can be absorbed in the single-particle energies since it is the same 
constant for the $\al=\pm 1$ components. This allows us to the treat the 
in-plane and out-of-plane cases at equal footing, assuming only that the 
Zeeman shift is large enough to eliminate the $\al=0$ component at the 
relevant temperatures.

The ensemble averages are calculated as corresponding 
time averages such that the system is allowed to thermalize for period 
$50\times 2\pi/\omega_{\perp}^{}$ and the time average is computed from $1250$ 
equally spaced samples. The sampling interval is 
$50\times 2\pi/\omega_{\perp}^{}$ ($100\times 2\pi/\omega_{\perp}^{}$) for the 
in-plane (out-of-plane) nematic phase. The randomized initial states are taken 
from the polar phase corresponding to zero magnetization. Otherwise the 
numerical implementation follows the description of Refs.~\cite{Blakie:2008,
Blakie:2008b}. In the HFP calculation for the in-plane nematic, we assume that 
there are no thermal atoms in the $\alpha=0$ component.

We keep the cutoff energy 
fixed, which causes the total number of atoms $N_{\mathrm{tot}}^{}$ to 
increase with increasing temperature. To accommodate to the varying atom 
number, we scale the temperature by the 
critical temperature $T_{0}^{}$ of a quasi-2D ideal Bose gas corresponding to 
the same total particle number~\cite{Bisset:2009}. For the in-plane 
(out-of-plane) nematic phase, $T_{0}^{}$ corresponds to the critical 
temperature of two (three) independent ideal Bose gases. For the ideal Bose 
gas, there is a standard relation between the temperature and the occupation 
number of the thermal component~\cite{Bisset:2009,Pitaevskii:2003} 
\be
N(T) = \sum_{n\neq 0}\,\frac{1}{e^{(\varepsilon_n^{}-\varepsilon_0^{})/\kb T}
-1},
\ee
where $\varepsilon_n^{}$ are the 3D harmonic oscillator energies. 
We calculate $T_0^{}$ numerically by solving the equation $N(T_0^{}) = 
\gamma N_{\mathrm{tot}}$ where $\gamma = 1/2$ for the in-plane phase and 
$\gamma=1/3$ for the out-of-plane phase. We denote the reduced temperatures 
by $T'$ and the bare temperatures by $T$. In experiments, a precise control of 
the total atom number at different temperatures is difficult and hence the 
approach presented here is likely to describe well the realistic experimental 
conditions.

\section{Condensate and quasicondensate}  
The existence and the nature of the condensate and the quasicondensate 
components in antiferromagnetic Bose gases are particularly interesting due to 
the possibility of a fragmented condensate at zero 
temperature~\cite{Ho:2000}. Since the fragmentation in this case corresponds 
to the condensation of composite bosons to the $|\bm{k}=0\rangle$ state in 
the momentum space, it seems that also the fragmented condensate is destroyed 
by the thermal fluctuations in a homogeneous 2D system. In addition, the 
thermally activated HQVs render the single-mode approximation used in 
Refs.~\cite{Ho:2000,Mueller:2006} invalid and it is a 
nontrivial question whether the fragmented condensate can exist in 2D at 
finite temperatures. In this work, the presence of a significant thermal 
component renders a direct comparison to the zero temperature 
single mode calculations difficult.

In the homogeneous 2D case, algebraic order is expected in 
the paired state corresponding to $\Theta = \hat{\Phi}_0^{}\hat{\Phi}_0^{} 
- 2\hat{\Phi}_{+1}^{}\hat{\Phi}_{-1}^{}$~\cite{Mukerjee:2006} and 
inspection of the correlation function $\la \Theta^{\da}_{}(\br'_{\perp})
\Theta(\br_{\perp}^{})\ra$ could shed light on the superfluid properties of 
the antiferromagnetic spin-$1$ Bose gases. In this work, we are interested 
in the existence and the nature of a condensed component in spin-$1$ 
superfluids and consider therefore the one-body density matrix  
$\rho^{(1)}_{}(\bm{r}\alpha;\bm{r}'\beta) = \langle 
\hat{\Phi}_{\beta}^{\dagger}(\bm{r}')\hat{\Phi}_{\alpha}^{}(\bm{r})\rangle$ 
which can be sampled using the time averaging. Under the previous assumptions 
it separates into two parts containing the c-field part and the incoherent 
part. At low temperatures, we find that $\rho^{(1)}_{}$ has only a single 
macroscopic eigenvalue $N_0^{}$ and we refer to $N_0^{}/N_{\mathrm{tot}}^{}$ 
by the generic name ``condensate fraction.''

Above the critical temperature of condensation, $\rho^{(1)}_{}$ has several 
large eigenvalues although their fraction of $N_{\mathrm{tot}}^{}$ becomes 
vanishingly small. This thermally induced fragmentation~\cite{Mueller:2006} 
is, however, different from the fragmentation due to the ordering in the spin 
sector. Our results seem to be consistent with the idea of a hierarchy of 
transition temperatures such that the formation of a coherent condensate is 
followed by ordering in the spin sector, leading potentially to a fragmented 
condensate in the zero-temperature limit~\cite{Mueller:2006}. The condensate 
fraction is shown in Fig.~\ref{condensation} as a function of the reduced 
temperature $T' = T/T_0^{}$.

\begin{figure}[h!]
\centering
\includegraphics[width=0.475\textwidth]{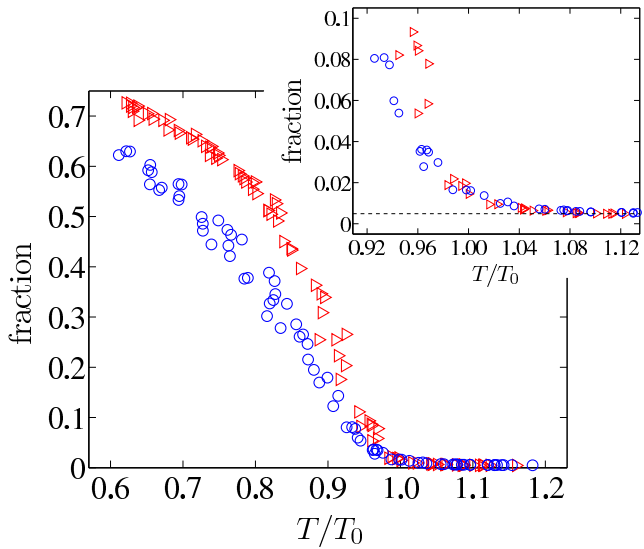}
\caption{\label{coh_cond} (Color online) The largest eigenvalue 
of the one-body density matrix ($N_0^{}$) normalized by the number of the 
c-field atoms ($\nc$) as a function of the reduced temperature. The inset 
shows the same quantity zoomed to the temperatures corresponding to the 
onset of a large eigenvalue. The out-of-plane nematic is denoted by (red) 
triangles, and (blue) circles correspond to the in-plane nematic phase. The 
dashed line is a guide for the eye.}
\end{figure}

For the scalar Bose gas, the quasicondensate component can be defined using 
the correlation function $\mathcal{C}(\br_{\perp}^{}) = 2\langle
\hat{\phi}^{\dagger}(\br_{\perp}^{})\hat{\phi}(\br_{\perp}^{})\rangle^2 - 
\langle[\hat{\phi}^{\dagger}(\br_{\perp}^{})\hat{\phi}(\br_{\perp}^{})]^2
\rangle$~\cite{Prokofev:2001,Bisset:2009}, describing the 
part of the system with reduced total density fluctuations. 
In the spinor case, we may consider an equivalent quantity   
$\mathcal{C}(\br_{\perp}^{})= 2\langle\hat{\Phi}^{\dagger}_\al(\br_{\perp}^{})
\hat{\Phi}_\al^{}(\br_{\perp}^{})\rangle^2 - 
\langle[\hat{\Phi}^{\dagger}_\al(\br_{\perp}^{})
\hat{\Phi}_\al^{}(\br_{\perp}^{})]^2\rangle$ where the $z$ dependence in the 
expectation values is integrated out.  In the spin-$1$ case analyzed here, we 
observe within the HFP approximation that the fraction of the component with 
suppressed total density fluctuations given by $\int\dr_{\perp}^{}\,
\sqrt{\mathcal{C}(\br_{\perp}^{})}/N_{\mathrm{tot}}^{}$,   
decreases very slowly with increasing temperature. This illustrates the role 
of the thermally induced intercomponent density fluctuations. 

We determine the quasicondensate component by considering the total density 
fluctuations restricted to the c-field region and define the quasicondensate 
density as 
\begin{equation}
n_{qc}^{}(\bm{r}) = \sqrt{2\langle |\psiic(\bm{r})|^2 \rangle^2-\langle 
|\psiic(\bm{r})|^4 \rangle}.
\end{equation}
For the parameters discussed in Section~\ref{defects}, the c field is of the 
form $\psiic(\br) = \psiic(\br_{\perp}^{})\varphi_0^{}(z)$, where 
$\varphi_0^{}(z)$ is the lowest harmonic oscillator state. Hence, the 
$z$ dependence in $n_{qc}^{}(\bm{r})$ can be integrated out.
The quasicondensate fraction is shown in Fig.~\ref{condensation} and it 
persists at the temperatures where the condensate fraction becomes negligible. 
The critical temperature for the formation of the coherent condensate as well 
as the onset of the quasicondensate are the same for the in-plane and the 
out-of-plane nematic phases, and they take place at temperatures 
$T' = 0.97\pm 0.02$ and $T' = 1.16\pm 0.04$, respectively 
(from Fig.~\ref{condensation}). In addition, the quasicondensate fraction is 
essentially the same at equal temperatures in both cases. Due to the reduced 
total density fluctuations at all temperatures, the quasi-condensate 
component is delocalized to the entire spatial extent of the c-field atoms 
whereas the condensate component tends to be localized to the region where 
HQVs and skyrmions are rare (see Fig.~\ref{densities}).

Since the reduced temperature for the onset of a condensate is the same for 
both nematic phases within the numerical accuracy, it is natural to ask if it 
is caused by the condensate depletion due to the incoherent atoms. Although 
the number of the incoherent region atoms is large near the onset of the 
condensate, the same onset temperature for the condensate is found if only 
the c-field atoms are taken into account. In Fig.~\ref{coh_cond}, the fraction 
$N_0^{}/\nc$ is shown as a function of the reduced temperature, indicating 
that the onset of a large eigenvalue takes place at equal temperatures for 
both nematic phases. Hence, the onset of a nonzero condensate fraction at the 
same temperature for both nematic phases does not depend on the depletion of 
the condensate due to the incoherent region atoms.

\section{Proliferation of topological defects} 
\begin{figure*}[t!]
\centering
\includegraphics[width=0.975\textwidth]{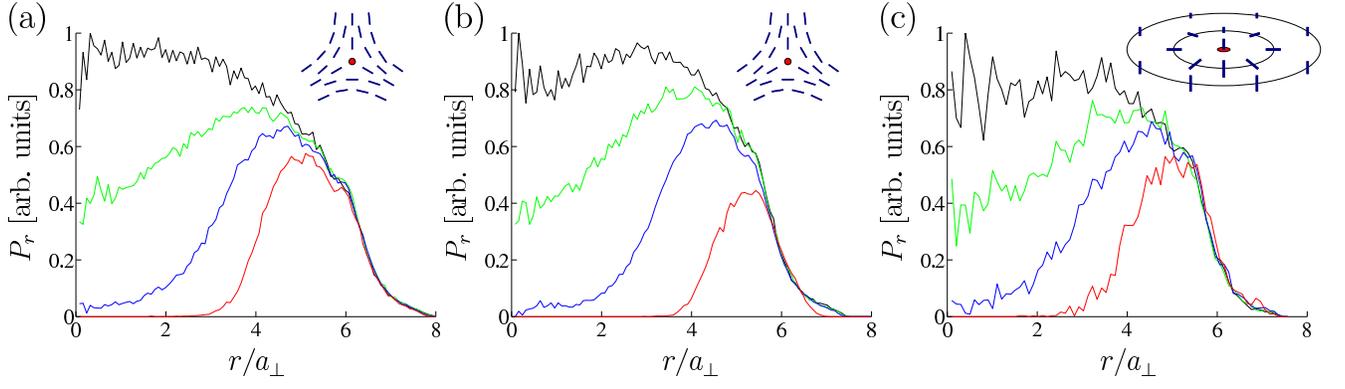}
\caption{\label{probabilities} (Color online) Radial probability density 
 for detecting (a) a HQV in the in-plane nematic, (b) a HQV in the 
 out-of-plane nematic, and (c) a skyrmion in the 
 out-of-plane nematic phase. The reduced temperatures are given by (from top 
 to  bottom in each panel)  $1.05,\,0.95,\,0.82,\,0.63$ in (a), 
 $1.00,\,0.94,\,0.89,\,0.64$ in (b), and $1.00,\,0.96,\,0.92,\,0.84$ in (c).}
\end{figure*}

In trapped atomic gases, the 
characteristic feature of the crossover from a BKT type of superfluid to a 
normal fluid is a proliferation and a subsequent propagation of free vortices 
from the edge of the cloud to the central region of the trap. 
Since HQVs are nonsingular defects, they persist at the edge of the cloud to 
relatively low temperatures (Fig.~\ref{densities}) and the system can be 
considered to have concentric shells of normal fluid and BKT superfluid with 
the center of the trap occupied by the condensate.  
We analyze the BKT crossover by studying the HQV occupation probability 
density $P_r^{}$~\cite{Simula:2008}. An estimate 
for the crossover temperature is  obtained from the temperature at which 
$P_r^{}$ becomes nonzero near the center of the trap (see 
Fig.~\ref{probabilities}). From this analysis, the BKT crossover takes 
place roughly at the reduced temperature $T_{\scriptscriptstyle \mathrm{BKT}}'
=0.82\pm 0.05$ for the in-plane nematic and at 
$T_{\scriptscriptstyle \mathrm{BKT}}'= 0.89\pm 0.04$ for the 
out-of-plane nematic phase. 

Since the reduced BKT crossover temperature obtained previously has a rather 
large uncertainty, we cannot conclusively determine the relation between the 
crossover temperatures for the two nematic phases. However, since the order 
parameter has a different symmetry in the in-plane and out-of-plane phases, 
there is no reason to assume that the crossover temperature is the same. 
For the in-plane nematic phase the symmetry is reduced to 
$[U(1)\times S^1]/\mathbbm{Z}_2$ while in the case of an out-of-plane nematic 
it is $[U(1)\times S^2]/\mathbbm{Z}_2$, allowing the existence of skyrmions 
which in the homogeneous case render the system 
spin-disordered. In a finite-size system, the thermal activation of 
skyrmions depends on the characteristic size of skyrmions compared to that of  
the system, and we find that skyrmions start 
to appear only at relatively high temperatures near the BKT crossover,  
see Fig.~\ref{probabilities}.  

The effect of skyrmions to the crossover temperature 
$T_{\scriptscriptstyle \mathrm{BKT}}$ can be illustrated by considering 
the statistical probability for the activation of a skyrmion or a pair of 
HQVs. The probability is proportional to the Boltzmann factor 
$\exp(-\Delta\mathcal{F}/\kb T)$ 
where $\Delta\mathcal{F} = \Delta\mathcal{E} - T\Delta S$ is the free-energy 
change associated with the creation of a given defect. The critical 
temperature for the activation of different defects can be estimated from 
the condition $\Delta\mathcal{F}=0$. In a uniform 2D system, the entropy 
change associated with the creation of a skyrmion can be approximately 
evaluated as~\cite{Kosterlitz:1972,Kosterlitz:1973,Simula:2005} 
$\Delta S_{\mathrm{sk}}^{} = 2\kb\ln(\ell/r_0^{})$ where $\ell$ is the linear 
size of the system and $r_0^{}$ the characteristic size of a skyrmion. 
The skyrmion energy $\mathcal{E}_{\mathrm{sk}}^{} = 4\pi\hbar^2\varrho/m$ is 
finite and independent of the size of the skyrmion (see Appendix). 
Hence, the free energy is always negative for a large enough system and 
skyrmions exist at all temperatures in the thermodynamical 
limit~\cite{Belavin:1975}. 

Since the BKT transition is driven by the dissociation of bound pairs of 
vortices and antivortices~\cite{Berker:1978}, we estimate the critical 
temperature using the Boltzmann probability for the appearance of free 
vortices~\cite{Kosterlitz:1972,Kosterlitz:1973}. In the Appendix, the energy 
of a free HQV is shown to be $\mathcal{E}_{\mathrm{HQV}}^{} = \frac{\pi\hbar^2
\varrho}{2m}\ln(\ell/r_1^{})$, where $r_1^{}$ is the size of the HQV core. The 
corresponding entropy change is $\Delta S_{\mathrm{HQV}}^{} = 
2\kb\ln(\eta\ell/r_1^{})$, where $\eta < 1$ in the presence of skyrmions due 
to screening. This would result in a higher critical temperature for the 
activation of free HQVs in the out-of-plane nematic phase where skyrmions are 
allowed. In nonuniform finite-size systems, the mechanism is different since 
the skyrmions appear only near $T_{\scriptscriptstyle \mathrm{BKT}}$. 
The thermal fluctuations generate skyrmions first at the boundary of the cloud 
and since the skyrmion energy is independent of its size, this process is not 
strongly affected by the pre-existing HQVs. The generation of skyrmions can 
prevent the thermal fluctuations from breaking the HQV--anti-HQV pairs, 
thereby giving rise to the higher crossover temperature.

\begin{figure}[h!]
\centering
\includegraphics[width=0.42\textwidth]{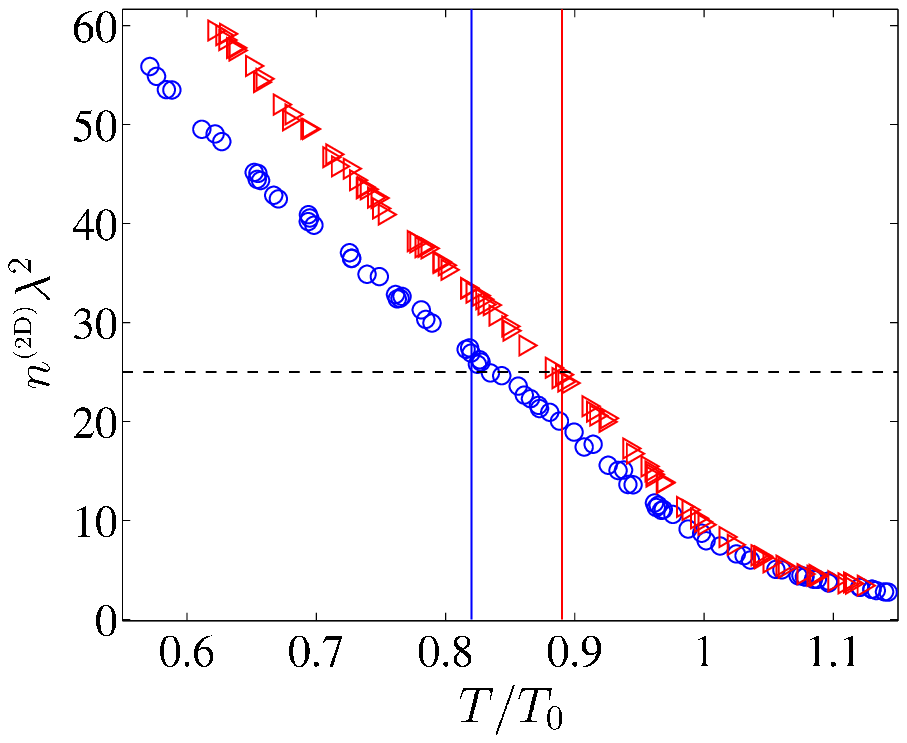}
\caption{\label{bkt_trans} (Color online) The phase-space density 
$\bar{n}_c^{\scriptscriptstyle\mathrm{(2D)}}\lambda^2$ as function of the  
reduced temperature for the in-plane nematic (blue circles) and out-of-plane 
nematic (red triangles). The solid lines denote the BKT crossover temperatures 
$T_{\scriptscriptstyle \mathrm{BKT}}'= 0.82$ and 
$T_{\scriptscriptstyle \mathrm{BKT}}'= 0.89$ for the in-plane and 
the out-of-plane nematic phases, respectively. The dashed line corresponds to 
the phase-space density 
$\bar{n}_{\mathrm{crit}}^{\scriptscriptstyle \mathrm{(2D)}}\lambda^2 = 25$.}
\end{figure}

The crossover temperature can also be studied using the 2D phase-space 
density $\bar{n}_c^{\scriptscriptstyle\mathrm{(2D)}}\lambda^2$, where 
$\bar{n}_c^{\scriptscriptstyle\mathrm{(2D)}}$ is the average 2D total density 
of the c-field atoms at $r_{\perp}^{}=0$ and $\lambda = \sqrt{2\pi\hbar^2/
m\kb T}$. We observe that the phase-space density takes roughly the 
value $25$ for both nematic phases at the respective reduced crossover 
temperatures (see Fig.~\ref{bkt_trans}). This result is to be contrasted with 
the single-component case where the transition to the superfluid phase takes 
place when the phase-space density is larger than the critical value 
\begin{equation}
\label{scalar_cond}
\bar{n}_{\mathrm{crit}}^{\scriptscriptstyle\mathrm{(2D)}}\lambda^2 = 
\ln(C/\tilde{g}),
\end{equation} 
where $\tilde{g} = \sqrt{8\pi}\,a/a_z^{}$, $a$ is the $s$-wave scattering 
length in the single component system, and $C\approx380$~\cite{Prokofev:2001,
Bisset:2009}. For the spin-$1$ system considered here, the corresponding 2D 
coupling constants can be defined as 
$\tilde{c}_0^{} = \sqrt{8\pi}(a_0^{}+2a_2^{})/3a_z$ and $\tilde{c}_2^{} = 
\sqrt{8\pi}(a_2^{}-a_0^{})/3a_z$. The parameters defined in 
Sec.~\ref{formalism} yield $\tilde{c}_0^{} = 0.028$ and $\tilde{c}_2^{} = 
0.0011$. The value of $\tilde{c}_0^{}$ is comparable to the experimental 
value $\tilde{g} = 0.02$ in the single-component system of  
Ref.~\cite{Clade:2009} where the spin-dependent interaction is absent.
A simple-minded application of the scalar case condition~\eqref{scalar_cond} 
using either of the coupling constants $\tilde{c}_0^{}$ and $\tilde{c}_2^{}$ 
with $C=380$ yields much lower values than $\bar{n}_c^{\scriptscriptstyle
\mathrm{(2D)}}\lambda^2 = 25$. This suggests that if a condition analogous to 
Eq.~\eqref{scalar_cond} exists for the spinor case, its form is different 
from~\eqref{scalar_cond}.

An important check for the observed crossover temperatures is the superfluid 
density which is predicted to change in the spinor case noncontinuously 
across the phase transition by the amount~\cite{Mukerjee:2006} 
\be
\label{bkt_cond}
\Delta\rho_s^{} = 8m^2\kb
T_{\scriptscriptstyle \mathrm{BKT}}^{}/(\pi\hbar^2).
\ee
Hence, the universal jump in the superfluid density is four times larger 
than in the single-component case. To use this property to check the 
consistency of the crossover temperatures, an independent computation of the 
superfluid density is required. Since the system is inhomogeneous, the central 
part of the system is typically in the superfluid state while the outer part 
corresponds to normal fluid. This renders methods such as the computation of 
the helicity modulus~\cite{Fisher:1973,Mukerjee:2006} inapplicable since they 
require a uniform system without coexisting phases. 

The HQVs are nonsingular vortices and the nonclassical moment of 
inertia~\cite{Holzmann:2008} does not capture the BKT crossover. 
The phenomenological models for the trapped systems in the single-component 
case~\cite{Holzmann:2008,Bisset:2009} make use of the 
condition~\eqref{scalar_cond} and assume explicitly a sudden 
change in the superfluid density by the universal value $2m^2\kb
T_{\scriptscriptstyle \mathrm{BKT}}^{}/(\pi\hbar^2)$.
Hence, there is a clear incentive for further investigations of the 
superfluid properties of spinor Bose gases, in particular for the 
determination of the superfluid fraction without making use of 
Eq.~\eqref{bkt_cond}. We note that the condition 
$\bar{n}_{\mathrm{crit}}^{\scriptscriptstyle \mathrm{(2D)}}\lambda^2 = 25$ 
is consistent with Eq.~\eqref{bkt_cond} since the crossover temperature 
yields ($\rho_s^{} = mn_s^{}$) $n_s^{}/\bar{n}_{\mathrm{crit}}^{
\scriptscriptstyle \mathrm{(2D)}} = 0.64$. Using the scalar 
condition~\eqref{scalar_cond} with $\tilde{g}$ equal to either 
$\tilde{c}_0^{}$ or $\tilde{c}_2^{}$ gives $n_s^{}/\bar{n}_{\mathrm{crit}}^{
\scriptscriptstyle \mathrm{(2D)}} > 1$, indicating that the scalar 
condition~\eqref{scalar_cond} is not valid for the spinor case.

We note that the uncertainty in the determination of the BKT 
crossover temperature in terms of the reduced temperature $T_{
\scriptscriptstyle \mathrm{BKT}}^{}/T_0$ allows in principle the crossover 
temperature to be the same for both nematic phases. However, 
if there is a critical value for the phase-space density 
independent of the type of the nematic ordering, then the data in 
Fig.~\ref{bkt_trans} yield different reduced crossover temperatures if 
they are below the condensation temperature $T'\approx 0.97$. 
We also note that it is numerically difficult to distinguish 
between skyrmions and merons when there are large fluctuations in the 
direction of the magnetic axis $\hat{\bm{n}}(\bm{r}_{\perp}^{})$, but in an 
analogy to the homogeneous 2D situation, we refer to these out-of-plane 
defects as skyrmions. In the in-plane case, skyrmions are not allowed but, 
instead, integer vortices corresponding to winding $2\pi$ of in the magnetic 
axis around the vortex core can take place. Such vortices seem to remain 
suppressed suggesting that they are irrelevant for the BKT crossover. 

\section{Discussion} 
We have analyzed the realization of the 
BKT transition in antiferromagnetic spin-$1$ Bose gases under typical 
experimental conditions. We have found a hierarchy of crossover temperatures 
corresponding to the onset of a quasicondensate at a high temperature and the 
formation of a coherent condensate at a lower temperature, followed by a 
BKT-type of crossover to a superfluid state as the temperature decreases.  
The investigation of the probability density for the excitation of skyrmions 
and half-quantum vortices did not unequivocally determine the relation between 
nematic ordering and the magnitude of the crossover temperature. The 
subsequent inspection of the 2D phase-space density at the center of the trap 
suggested that the crossover temperature expressed in terms of the reduced 
temperature could be slightly higher for the out-of-plane phase. Further 
investigations are still needed to confirm this scenario.
The finite size of the system is manifested as a 
finite activation temperature for the skyrmions and the thermal fluctuations 
start to generate skyrmions only near the crossover temperature. It 
remains an open question if another crossover to a fragmented condensate 
takes place in the zero-temperature limit. 

We expect that the fractional population of 
different Zeeman sublevels can be controlled using rf pulses and 
magnetic-field gradients~\cite{Vengalattore:2009} to allow the experimental 
realization of the in-plane and the out-of-plane nematic phases. Using the 
time-of-flight imaging combined to the Stern-Gerlach separation of the 
different Zeeman sublevels, formation of the condensate component can be 
observed. The ferromagnetic cores of the HQVs could be detected by 
imaging the magnetization of the gas~\cite{Higbie:2005,Vengalattore:2009} and  
the same technique can in principle be extended to image directly also the 
spin quadrupole order~\cite{Carusotto:2004,Higbie:2005,Saito:2008b}. 
Interference experiments~\cite{Hadzibabic:2006} can also be useful to 
demonstrate the existence of free vortices around the BKT crossover 
temperature.

\acknowledgments

The authors acknowledge the Jenny and Antti Wihuri Foundation, the Emil 
Aaltonen Foundation, the Japan Society for the Promotion of Science (JSPS), 
and the Academy of Finland for financial support and the Center for Scientific 
Computing Finland (CSC) for computing resources. 

Part of this research was carried out in the Centre for Quantum Computer 
Technology supported by the Australian Research Council, the Australian 
Government, the U.S.~National Security Agency (NSA), and the 
U.S.~Army Research Office (ARO) (under Contract No.~W911NF-08-1-0527).

\appendix*
\section{Energies of skyrmions and HQV--anti-HQV pairs in uniform systems}

The skyrmion configuration can be represented on a Cartesian basis 
such that $\vec{\Psi} = \sqrt{\varrho}e^{i\theta}\hat{\bm{n}}$, with
\be
\label{skyrm}
\hat{\bm{n}} = \big(\sin\beta(\rho)\cos\varphi,\sin\beta(\rho)
\sin\varphi,\cos\beta(\rho)\big),
\ee
where $(\rho,\varphi)$ denote the polar coordinates and function 
$\beta(\rho)$ satisfies the boundary conditions $\beta(0) = 0$ and 
$\beta(\rho) = \pi$ for $\rho > r_0^{}$. A meron (half-skyrmion) is obtained 
with $\beta(\rho) = \pi/2$ for $\rho > r_0^{}$. We assume a uniform system 
such that the density $\varrho$ is a constant for skyrmions and HQVs. For the 
skyrmion configuration~\eqref{skyrm}, the $U(1)$ phase $\theta$ can be taken 
to be a constant.

The low-energy theory for the polar phase is the nonlinear $\sigma$ 
model (\nlsm) of the form~\cite{Mukerjee:2006} 
\be
\label{non-linear}
\mathcal{L} = 
\frac{K}{2}\int\mathrm{d}\br_{\perp}^{}\,[(\nabla\hat{\bm{n}})^2 + 
(\nabla\theta)^2],
\ee
where the superfluid stiffness is $K=\hbar^2\varrho/m$. The \nlsm~has a 
conformal invariance such that the energy of a skyrmion can become independent 
of the size $r_0^{}$. Furthermore, all configurations of the 
form~\eqref{skyrm} satisfy the condition $\mathcal{E}_{\mathrm{sk}}^{} \geq 
4\pi K$~\cite{Belavin:1975}. Hence we can take the energy of the skyrmion to 
be $\mathcal{E}_{\mathrm{sk}}^{} = 4\pi K$.

Outside the vortex core for $\rho > r_1^{}$, the HQV configuration corresponds 
to $\theta=\varphi/2$ and 
\be
\label{hqv}
\hat{\bm{n}} = (\cos\frac{\varphi}{2},\sin\frac{\varphi}{2},0).
\ee
Assuming that the systems has linear size $\ell$, substitution of~\eqref{hqv} 
into~\eqref{non-linear} yields the energy 
$\mathcal{E}_{\mathrm{HQV}}^{} = \frac{\pi K}{2}\ln(\ell/r_1^{})$. The HQV 
energy does not include the contribution 
from the vortex core which is negligible in the thermodynamical limit.
Furthermore, the usual arguments~\cite{Pitaevskii:2003} can be used to 
conclude that the energy of a HQV--anti-HQV pair is in the leading order 
$\mathcal{E}_{\mathrm{pHQV}}^{} = \pi K \ln(d/r_1^{})$, where $d$ is 
the distance between the vortex cores. 

\bibliography{manu}

\end{document}